\documentclass[12pt]{article}
\usepackage{a4}
\usepackage{amsfonts}
\usepackage[dvips]{graphicx}
\begin{document}
\begin{titlepage}
\vskip0.5cm
\begin{flushright}
DESY 01-103\\
\end{flushright}
\vskip0.5cm
\begin{center}
{\Large\bf Speeding up the Hybrid-Monte-Carlo algorithm}
\vskip 0.3cm
{\Large\bf for dynamical fermions}
\vskip 0.3cm
\end{center}
\vskip 1.3cm
\centerline{
Martin Hasenbusch
}
\vskip 0.4cm
\centerline{\sl  NIC/DESY-Zeuthen}
\centerline{\sl  Platanenallee 6, D-15738, Germany}
\vskip 0.3cm
\centerline{e-mail: Martin.Hasenbusch@desy.de}

\vskip 0.4cm
\begin{abstract}
We propose a modification of the Hybrid-Monte-Carlo algorithm that allows for 
a larger step-size of the integration scheme at constant acceptance rate.
The key ingredient is 
that the pseudo-fermion action is split into two parts. We test our proposal 
at the example of the two-dimensional lattice Schwinger model with two 
degenerate flavours of
Wilson-fermions.
\end{abstract}
\end{titlepage}
\section{Introduction}
The Hybrid-Monte-Carlo algorithm \cite{hybrid} has become the standard 
algorithm to simulate lattice QCD with dynamical fermions. E.g. it has 
been used in the recent large scale study \cite{CP-PACS} of the spectrum
of light hadrons.

At present, most simulations are performed for two flavours of mass-degenerate
sea-quarks, where the mass is by 
far larger than the masses of the up and down quark. An elaborate 
extrapolation of the data towards the chiral limit is needed.

Therefore it is desirable to further reduce the mass of the sea-quarks.
However, reducing the mass of the sea-quarks,
i.e. approaching $\kappa_c$ in the case of Wilson fermions, 
the cost of the simulation increases for at least three reasons 
(see e.g. ref. \cite{QCDwith}):

\begin{itemize}
\item
The condition number of the Dirac-matrix  
increases. Hence the number of iterations that are needed for the 
inversion of the Dirac-matrix increases.

\item
The autocorrelation times in units of trajectories 
increase as $\kappa_c$ is approached.

\item
 The step-size of the % leap-frog 
 integration scheme has to be decreased as 
$\kappa_c$ is approached to maintain a constant acceptance rate. This means
that for one trajectory more inversions of the Dirac matrix have to be 
performed. 
This effect can be nicely seen in table II of ref. \cite{CP-PACS}. 

\end{itemize}

While much work has been devoted to deal with the first two problems,
little attention has been paid to the third.
Recently it has been noticed that  the step-size at a given 
acceptance rate depends on the precise form of the 
pseudo-fermion action \cite{Pe,alphabench}.  It was shown that 
preconditioning of the fermion matrix leads to larger step-sizes 
in the Hybrid-Monte-Carlo algorithm.

In this paper we propose a rather simple modification of the 
pseudo-fermion action that allows to increase the step-size at fixed
acceptance rate. In particular, we split the pseudo-fermion action
into two parts, separating (partially) the small and the large eigenvalues
of the Dirac-matrix. 

The paper is organised as follows. In section 2 we introduce the modified 
pseudo-fermion action. We test our proposal at the example of 
the two-dimensional two-flavour Schwinger model. In particular, we show that
our idea can be applied in addition to even-odd preconditioning that is 
discussed in section 3.
In section 4 we present our numerical results. Finally we give our conclusions 
and an outlook.

\section{The modified pseudo-fermion action}
The following discussion is rather general and applies, e.g., to 
four-dimensional lattice QCD with Wilson fermions as well as the 
two-dimensional Schwinger model on the lattice.
We start our discussion with 
the partition function for two degenerate flavours of Wilson 
fermions: 
\begin{equation}
Z = \int \mbox{D}[U] \exp(-S_G(U)) \; \; \mbox{det} M(U)^\dag M(U) \;\;,
\end{equation}
where $U$ is the gauge field, $S_G(U)$ the gauge action and 
\begin{equation}
M(U)=1 - \kappa H(U) \;\;
\end{equation}
the fermion matrix, where $\kappa$ is the hopping parameter.

In the Hybrid-Monte-Carlo algorithm, the determinant 
$\mbox{det} M^\dag M $ is represented by the integral over an auxiliary 
field (pseudo-fermions) $\phi$:
\begin{equation}
\mbox{det} M^\dag M \; \propto \; \int \mbox{D}[\phi] \mbox{D}[\phi^\dag] \;
\exp(-|M^{-1} \phi|^2) \;\;.
\end{equation}
Hence the action, as a function of the gauge-field and the 
pseudo-fermion fields,
is given by
\begin{equation}
S(U,\phi)\; = \;S_G(U)\; +\; S_F(U,\phi)
\end{equation}
with
\begin{equation}
\label{fermionaction}
S_F(U,\phi)
\;=\; |M(U)^{-1} \phi|^2 \;\;.
\end{equation}

Our new idea is to split the fermion matrix into two factors. Each  factor
is represented by an integral over an auxiliary field:
\begin{equation}
\label{splitting}
\mbox{det} M^\dag M \; \propto \;\int \mbox{D}[\psi] \mbox{D}[\psi^\dag] \;
\mbox{D}[\phi] \mbox{D}[\phi^\dag] \;
\exp(-|\tilde M^{-1}  \psi|^2) \;
\exp(-|\tilde M M^{-1} \phi|^2) \;\;,
\end{equation}
where the auxiliary fermion matrix $\tilde M$ is chosen as
\begin{equation}
\tilde M=1 - \tilde \kappa H \;\;
\end{equation}
with $ \tilde \kappa < \kappa$. I.e. the pseudo-fermion action is now
given by the sum of the two terms
\begin{equation}
\label{modif1}
S_{F1}(U,\psi)
\;=\; |\tilde M(U)^{-1} \psi|^2
\end{equation}
and
\begin{equation}
\label{modif2}
S_{F2}(U,\phi)
\;=\; |\tilde M(U) M(U)^{-1} \phi|^2 \;\;.
\end{equation}
The condition number of $\tilde M$ as well as $R= M \tilde M^{-1}$ is reduced
compared with the original matrix $M$. This is the reason, why we expect,
similar to preconditioning \cite{Pe,alphabench},
that the step-size in the Hybrid-Monte-Carlo
can be increased. 
In the case of $\tilde M$ the condition number is reduced for typical 
gauge-configurations since
we have chosen $\tilde \kappa < \kappa$. Next we consider $R$.
Let us write
\begin{equation}
\tilde M = a M + b 
\end{equation}
with
\begin{equation}
\label{aeq}
a = \frac{\tilde \kappa}{\kappa} \;, \;\;\;\; b = 1-a \;\;\;.
\end{equation}
Hence
\begin{equation}
\label{Rminus1}
R^{-1}=a +b M^{-1} \;\;\;.
\end{equation}
For $\kappa$ close to $\kappa_c$, $|\lambda_{min}|^{-1} \gg 1$ and 
$|\lambda_{max}| = O(1)$, where $\lambda_{min}$ and $\lambda_{max}$ are
the minimal and maximal eigenvalues of $M$. Hence the condition number of
$R$ is essentially reduced by a factor of $b$ compared with the condition 
number of $M$.

In the Hybrid-Monte-Carlo, the variation of the action with respect to 
the gauge field has to be computed. In the case of the standard pseudo-fermion 
action~(\ref{fermionaction}) one obtains
\begin{eqnarray}
\label{variation}
 \delta S_{F} &=& - \phi^\dag 
 \left[M^{\dag -1} M^{-1} \;\delta M \; M^{-1}  +
       M^{\dag -1} \;\delta M^\dag  \; M^{\dag -1} M^{-1} \right] 
 \phi  \nonumber \\
&=&-\left[Y^\dag \;\delta M \;X + X^\dag \; \delta M^\dag \; Y \right] \;\;,
\end{eqnarray}
where $X= M^{-1} \phi$ and $Y = M^{\dag -1} X$.

In the case of the split pseudo-fermion action $\delta S_{F1}$ can be 
computed exactly as in eq.~(\ref{variation}). Also $\delta S_{F2}$  can  
easily be computed. 
\begin{eqnarray}
\label{variation2}
\delta S_{F2} &=& - \phi^\dag
 \left[R^{\dag -1} R^{-1} \;\delta R \; R^{-1}  +
       R^{\dag -1} \;\delta R^\dag  \; R^{\dag -1} R^{-1} \right]
 \phi  \nonumber \\  
  &=& - \phi^\dag                                              
 \left[(a+b M^{\dag -1})  M^{-1} \;\delta M \; M^{-1}  +
       M^{\dag -1} \;\delta M^\dag  \; M^{\dag -1} (a+b M^{-1})  \right]
 \phi  \nonumber \\                                                
&=&-\left[Y^\dag \;\delta M \;X + X^\dag \; \delta M^\dag \; Y \right] \;\;,
\end{eqnarray}
where we have used $R^{-1} \;\delta R\;R^{-1} = b \;M^{-1} \;\delta M\;M^{-1}$
which follows from eq.~(\ref{Rminus1}). The auxiliary vectors $X$ and $Y$
are modified compared with eq.~(\ref{variation}):
$X=M^{-1} \phi$ and $Y=M^{\dag -1} (a \phi + b X)$. 

I.e. the computational cost to compute
the variation of the second part of the modified 
pseudo-fermion action~(\ref{modif2}) is much the same as for the standard 
pseudo-fermion action~(\ref{fermionaction}). In both cases we have to 
apply $M^{-1}$ and $M^{\dag -1}$ to a vector.

Computing the variation of the first part of the modified 
pseudo-fermion action~(\ref{modif1}) means an overhead compared with the 
standard case. However, since $\tilde \kappa < \kappa$ fewer iterations 
are needed to compute $\tilde M^{-1} \psi$ than $M^{-1} \phi$.

We also should note that the modification can be very easily implemented, 
given a standard Hybrid-Monte-Carlo program is at hand. One just has to 
add a second pseudo-fermion field for $R$ and implement the modified 
definition of the auxiliary vectors $X$ and $Y$.

In the following simulations we like to test whether the modified 
pseudo-fermion action indeed allows for a larger step-size at constant
acceptance rate and whether this larger step-size outweighs the 
computational overhead discussed above.

\section{The Schwinger model}
We have tested our proposal to split the fermion matrix in the two
dimensional Schwinger model with two degenerate flavours of Wilson fermions.
It has been used frequently as a toy-model to study properties 
of Monte Carlo algorithms. 

The gauge action of the Schwinger model is given by
\begin{equation}
 S_G= -\beta \sum_{x} \mbox{Re} \; U_{plaq,x}\;\; ,
\end{equation}
where
\begin{equation}
U_{plaq,x} = U_{x,0} \;\; U_{x+(1,0),1} \;\; U_{x+(0,1),0}^*
\;\; U_{x,1}^* \;\;\; ,
\end{equation}
where the link variables $U_{x,\mu}$ are elements of $U(1)$ and $U_{x,\mu}^*$
is the complex conjugate of $U_{x,\mu}$. $x$ labels the 
points of the two-dimensional square lattice and $\mu$ gives the direction.
The lattice constant is set to $a=1$.
The fermion matrix can be written as
\begin{equation}
M=1 - \kappa H \;\;,
\end{equation}
where the hopping part of the fermion matrix is given by
\begin{equation}
H= \sum_{\mu} \left(\delta_{x-\hat \mu,y} \; (1+\gamma_{\mu}) \;
U_{x-\hat \mu,\mu}
 +\delta_{x+\hat \mu,y} \; (1-\gamma_{\mu}) \; U_{x,\mu}^* \;
\right) \;\;\;,
\end{equation}
where
$\hat \mu$ is unit vector in $\mu$ direction.
The $\gamma$-matrices in two dimensions are given by the 
Pauli-matrices $\sigma$.
%where in two dimensions we choose the  $\gamma$-matrices as
%\begin{equation}
%\gamma_1 =   \left( \begin{array}{cc}
%  0   &  1  \\
%  1   &  0
%\end{array} \right)
%\;\;\;\;\;\;\;\;\;
%\gamma_2 =   \left( \begin{array}{cc}
%1   &  0  \\
%0   & -1
%\end{array} \right)
%\;\; .
%\end{equation}
In our simulations we applied even-odd preconditioning. See e.g. ref. 
\cite{QCDwith}.
The sites of the lattice are decomposed in even and odd sites. Then
the fermion matrix can be written in the form
\begin{equation}
 M=  \left( \begin{array}{cc}
    1_{ee}   &  -\kappa H_{eo}  \\
              -\kappa H_{oe}   &  1_{oo} 
\end{array} \right)  \; \; \;,
\end{equation}
where $H_{eo}$ connects odd with even sites and $H_{oe}$ vice versa.
For the fermion determinant the identity
\begin{equation}
\mbox{det} M = \mbox{det} (1_{ee}-\kappa^2 H_{eo} H_{oe})
\end{equation}
holds.
Hence the original problem is reduced by half in
the dimension (of the fermion matrix).
The pseudo-fermion field $\phi$ which is used for the stochastic estimate
of the fermion determinant lives
only on even sites. The even-odd preconditioned fermion matrix is
given by 
\begin{equation}
\label{Mprecon}
M_{ee} = 1_{ee} - \kappa^2 \;\; H_{eo} \; H_{oe} \;\; .
\end{equation}
In the following we shall apply our proposed splitting~(\ref{splitting}) to
the even-odd preconditioned fermion matrix $M_{ee}$. We use
\begin{equation}
\tilde M_{ee} = 1_{ee} - \tilde \kappa^2 \;\; H_{eo} \; H_{oe} \;\; .  
\end{equation} 
I.e. in eq.~(\ref{aeq}) $\kappa$ and $\tilde \kappa$ have to be replaced
by $\kappa^2$ and $\tilde \kappa^2$.

\section{Numerical results}
In our simulations with the Hybrid-Monte-Carlo algorithm we have 
used the standard leap-frog integration scheme. In order to separate 
the effects of the gauge action and the pseudo-fermion action  we applied
the two step-size method of Sexton and Weingarten \cite{SeWe}. In particular, 
we have chosen $n=4$ as ratio of the step sizes.  
The length of the trajectory was chosen randomly 
between $0.5$ and $1.5$.
We computed $M_{ee}^{-1} \phi$ with the BiCGstab algorithm. % \cite{BICG}.
In the rare cases, where BiCGstab did not converge, we used
the conjugate gradient algorithm. We started the iteration with 
$\phi$. As stopping criterion, we  
required that the residue is less that $10^{-10}$.

We simulated at parameters $\beta$ and $\kappa$ 
that were recently used by Peardon \cite{Pe} to study
the effects of preconditioning on the performance of the algorithm.

We performed a first set of simulations for $L=32$, $\beta=4.0$ and 
$\kappa=0.26$. 
In table \ref{L32table} we have summarised results of simulations using
various values $\tilde \kappa$. In all these simulations we generated 20000
trajectories. 1000 trajectories were discarded for thermalisation.
In all our simulations we
fixed the acceptance rate to about $0.8$. For this purpose we performed
a few preliminary simulations for each $\tilde \kappa$ with smaller statistics.

For $\tilde \kappa =0.0$ we obtained an acceptance rate of about $0.8$ with 
a step-size of ${\rm d}\tau=0.035$. Here ${\rm d}\tau$ is the step-size 
used for the pseudo-fermion action. The step-size for the gauge-action  
is ${\rm d}\tau/4$.
Increasing $\tilde \kappa$ the step-size can be 
increased up to about ${\rm d}\tau=0.07$ for $\tilde \kappa=0.22$. Increasing 
$\tilde \kappa$ to $0.255$ leads to a smaller step-size again. This 
increase is not surprising, since in the limit $\tilde \kappa \rightarrow
\kappa$ we recover the standard algorithm again.

The number of iterations $\tilde m$ and $m$ needed  to compute 
$\tilde M_{ee}^{-1} \psi$ and $M_{ee}^{-1} \phi$, respectively,
refer to the inversions performed for the accept/reject step 
at the end of the trajectory. I.e. they are not performed for equilibrium
configurations. Therefore it is not surprising that $m$ depends slightly 
on $\tilde \kappa$.

For $\tilde \kappa=0.22$ the numerical overhead to compute the modified 
fermion action compared with the standard one is rather moderate. The number
of iterations needed to compute $\tilde M_{ee}^{-1} \psi$ is only a small 
fraction of the iterations needed to compute $M_{ee}^{-1} \phi$.

We have measured  square Wilson-loops  up to size of $5\times 5$
and the topological charge, using the geometric definition.
The estimates of the expectation values were consistent among the simulations
that we have performed. Also, the results for the Wilson-loops of size 
$1\times 1$ and $4\times 4$ are consistent with those quoted by Peardon 
\cite{Pe} in his table 2.
It turned out that the auto-correlation times (in units of trajectories)
of the various observables
are, within statistical errors, independent of $\tilde \kappa$. 
Therefore, the runs can be 
compared solely on the ground of the numerical cost per trajectory.

The numerical costs are dominated by applications of $M_{ee}$ and $M_{ee}^\dag$.
Therefore, as a measure of the computational 
cost we took the number of applications of 
$M_{ee}$ and $M_{ee}^\dag$ per trajectory.
The minimum of costs is rather shallow and is located  around
$\tilde \kappa=0.22$. 
Compared with the standard simulation 
($\tilde \kappa=0$) we see a reduction of the cost by a factor of 
$24788/14898 \approx 1.66$.

\begin{table}
\caption{\sl \label{L32table}
Results for $L=32$, $\beta=4.0$ and $\kappa=0.26$. In the first column 
we give the hopping parameter $\tilde \kappa$ of the auxiliary fermion
matrix $\tilde M_{ee}$. The second column contains the step-size 
${\rm d} \tau$.
In the third and fourth column, the number of iterations needed to 
solve $\tilde M_{ee}^{-1}\phi $ and $M_{ee}^{-1}\psi $ is given, respectively.
In the fifth column we give the acceptance rate.
Finally, in the sixed column the cost, which is defined in the text, is
presented. Details are discussed in the text.
}
\begin{center}
\begin{tabular}{|l|l|l|l|l|l|}
\hline
%\rule[0mm]{0mm}{4mm}
\multicolumn{1}{|c}{$\tilde \kappa$}  & 
\multicolumn{1}{|c}{${\rm d}\tau$} & 
\multicolumn{1}{|c}{$\tilde m$} &
\multicolumn{1}{|c}{$m$} &
\multicolumn{1}{|c}{acc} &
\multicolumn{1}{|c|}{cost} \\
\hline
0.0  & 0.035 &                     & 203.42(21) & 0.808(4) & 24788(53) \\
0.18 & 0.056 &\phantom{0}14.576(5) & 204.85(16) & 0.793(4) & 17107(35) \\
0.20 & 0.063 &\phantom{0}19.771(5) & 205.42(13) & 0.803(3) & 15724(31) \\
0.21 & 0.067 &\phantom{0}23.724(6) & 205.59(13) & 0.784(4) & 15120(29) \\
0.22 & 0.07  &\phantom{0}29.440(8) & 206.08(13) & 0.784(4) & 14898(29) \\
0.23 & 0.072 &\phantom{0}38.394(11)& 206.26(13) & 0.784(3) & 15058(28) \\  
0.24 & 0.07  &\phantom{0}54.596(21)& 206.61(12) & 0.801(3) & 16515(33) \\
0.25 & 0.07  &\phantom{0}91.91(6)  & 206.84(16) & 0.794(3) & 18958(37) \\
0.255& 0.065 &          134.29(10) & 207.09(12) & 0.774(3) & 23181(47) \\
\hline
\end{tabular}
\end{center}
\end{table}

Finally we performed simulations for $L=64$, $\beta=4.0$ at 
$\kappa=0.257$ and $\kappa=0.2605$.  Following Peardon \cite{Pe}, the 
pseudo-meson masses at these values of $\kappa$ are $a m_P = 0.210(3)$ 
and $a m_P = 0.124(5)$, respectively. 

Here we performed only simulations for the standard
case $\tilde \kappa=0$ and  for one non-trivial value of $\tilde \kappa$.
We have chosen $\tilde \kappa=0.22$ for $\kappa=0.257$ and
$\tilde \kappa=0.23$  for $\kappa=0.2605$.
The simulations for the $L=32$ lattice have shown that 
the minimum in the computational cost 
is rather shallow in $\tilde \kappa$. 
Therefore, we expect that the computational 
cost at our values of $\tilde \kappa$ is rather close to the optimum.

\begin{table}
\caption{\sl \label{L64table}
Results for $L=64$, $\beta=4.0$ for the two values of  $\kappa=0.257$ and
$\kappa=0.2605$. 
For $\kappa=0.257$ we have performed 10000 trajectories for
each value of $\tilde \kappa$. For $\kappa=0.2605$ we have performed
6600 trajectories for $\tilde \kappa=0.0$ and
8400 trajectories for $\tilde \kappa=0.23$. The notation is the same as
in table 1.
}
\begin{center}
\begin{tabular}{|l|l|l|l|l|l|l|}
\hline
%\rule[0mm]{0mm}{4mm}
\multicolumn{1}{|c}{$\kappa$}  & 
\multicolumn{1}{|c}{$\tilde \kappa$}  & 
\multicolumn{1}{|c}{${\rm d} \tau$} & 
\multicolumn{1}{|c}{$\tilde m$} &
\multicolumn{1}{|c}{$m$} &
\multicolumn{1}{|c}{acc} &
\multicolumn{1}{|c|}{cost} \\
\hline
0.257 & 0.0  &  0.04  &           & 195.8(2) &   0.784(5) &  21669(71) \\
0.257 &0.22  &  0.054 & 30.210(10)& 202.4(2) &   0.783(5) &  18681(55) \\
\hline
0.2605& 0.0  &  0.022 &           & 384.4(8) & 0.792(7)   &  74550(330) \\
0.2605& 0.23 &  0.051 &  39.69(2) & 398.2(7) & 0.788(5)   &  37038(130) \\
\hline
\end{tabular}
\end{center}
\end{table}

First of all we notice that for $\tilde \kappa=0$ the step-size ${\rm d} \tau$ 
has to be decreased when $\kappa$ is increased.  
For $\tilde \kappa \ne 0$ the decrease of the step-size with 
increasing $\kappa$ is much smaller.
I.e. the performance gain that we achieve with the modified pseudo-fermion action 
increases as $\kappa_c$ is approached. For $\kappa=0.2605$ with 
$\tilde \kappa=0.23$ we get an improvement of a factor $74550/37038 \approx 2$.

\section{Conclusions and Outlook}
We have proposed to use a modified pseudo-fermion action in the 
Hybrid Monte Carlo simulation. This modification is based on a factorisation
of the fermion matrix. The modification can be easily implemented, given a
code for the standard Hybrid-Monte-Carlo algorithm is at hand. 
We have tested our proposal at the example of the 
two-dimensional two-flavour Schwinger model with Wilson fermions.
The numerical tests have shown that the modification of the fermion action 
indeed allows for a larger step-size of the leap-frog integration scheme at 
a fixed acceptance rate. Moreover, the increased step size outweighs 
the numerical
overhead due to the modified action. For our largest value of $\kappa$
we found a net gain in performance of a factor of two. 

There remain a number of open problems. First of all we would like to 
test our proposal for lattice QCD. %\cite{inprogress} 
We would like to study 
more systematically the dependence of the improvement on the parameters 
$\beta$, $\kappa$ and the lattice size. Also a better theoretical 
understanding of how the performance of the Hybrid-Monte-Carlo depends on 
the form of the pseudo-fermion action would be helpful. It would 
be interesting to test how higher-order integration schemes \cite{SeWe}
perform in combination with the modified pseudo-fermion action. Also, 
one might incorporate our new idea in the Polynomial-Hybrid-Monte-Carlo
algorithm \cite{PHMC1,PHMC2}.

\section{Acknowledgements}
I like to thank K. Jansen, R. Sommer and U. Wolff for discussions.

\end{document}